\documentclass[12pt]{iopart}

\usepackage{harvard}
\usepackage{iopams}
\usepackage[dvips]{color}
\usepackage{graphicx}

\begin{document}

\title[Diffusion of particles in an expanding sphere with an absorbing
boundary]{Diffusion of particles in an expanding sphere with an absorbing boundary}

\author{K. Forsberg$^1$ and A. R. Massih$^{2,3}$\footnote{Corresponding author.}}

\address{$^1$Brons\aa ldersg. 6, SE-723 51 V\"{a}ster\aa s, Sweden\\
$^2$Quantum Technologies, Uppsala Science Park, SE-751 83 Uppsala and \\
$^3$Malm\"{o} University, SE-205 06 Malm\"{o}, Sweden}
\ead{alma@quantumtech.se}

\begin{abstract}
 We study the problem of particles undergoing Brownian motion in an
expanding sphere whose surface is an absorbing boundary for the
particles. The problem is akin to that of the diffusion of
impurities in a grain of polycrystalline material undergoing grain
growth. We solve the time dependent diffusion equation for particles
in a $d$-dimensional expanding sphere to obtain the particle density
function (function of space and time). The radius of sphere expands as
$R=(Kt)^\alpha$, where $\alpha$ is the growth exponent, $K$ a constant
and $t$ the time. This allows the calculation of the survival rate or
the total number of particles per unit volume as a function of
time. We have obtained particular solutions exactly for the case where
$d=3$ and a parabolic expansion of the sphere. Asymptotic solutions
for the particle density when the sphere expansion rate is small
relative to particle diffusivity  and vice versa are derived.

\end{abstract}

\maketitle

\section{Introduction}
\label{sec:intro}

The kinetics of many physical systems can be described by
first-passage properties of stochastic systems \cite{Redner_2001}.
In particular, the problem of the survivability of a diffusing
particle, or Brownian walker, in a confined domain (``cage'') with
moving and absorbing boundary (``fence'') has been a subject of
several recent studies
\cite{Krapivsky_Redner_1996,Redner_2001,Bray_Smith_2007a,Bray_Smith_2007b}.

\citeasnoun{Krapivsky_Redner_1996} evaluated the survival probability
of a diffusing particle $S(t)$ as a function of time $t$ within a
1-dimensional expanding cage of length $L(t)$ with absorbing
boundaries by solving the standard diffusion equation (heat equation)
in one dimension. Approximate methods were devised in the limit of
slow (adiabatic approximation) and fast (free approximation) motion of
the absorbing boundary. More specifically,
\citeasnoun{Krapivsky_Redner_1996} considered $L(t)\sim t^\alpha$ with
$\alpha <1/2$ for adiabatic approximation and $\alpha >1/2$ for free
approximation. Moreover, the case of $\alpha=1/2$, corresponding to
marginally expanding cage, was theoretically assessed, see also
\citeasnoun{Redner_2001}.

\citeasnoun{Bray_Smith_2007a} calculated the exact asymptotic survival
probability of a 1-dimensional Brownian particle, initially confined
at a position $x \in (-L(t), L(t))$ in the presence of two moving and
absorbing boundaries by solving a Fokker-Planck equation for the
system in the limit $t \to \infty$. Subsequently, they
\cite{Bray_Smith_2007b} extended their method to treat the case of the
$d$-dimensional system and in particular $d=3$. Similarly, in the
latter work, they considered the steady-state case ($t \to \infty$) and
assumed the absorbing sphere's size evolves as $L(t)=L_0+vt$, where
$L_0$ is the initial size and $v$ is the constant velocity of the
boundary.

The aforementioned deliberations have relevance to the problem of
diffusing particles (impurities) in a grain of polycrystalline
material under grain growth \cite{Forsberg_Massih_2007}. Grain growth
is a kinetic process at which the mean grain size of an aggregate of
crystals increases during annealing at an elevated high temperature
\cite{Atkinson_1988}. In this course, the larger grains are apt to
increase in size at the expense of smaller grains which collapse,
causing the total number of grains to decrease.  Impurities in solid
solution, prevalent in polycrystalline materials, which usually
undergo diffusive motion at elevated temperatures, are in addition
subjected to a moving or stretching medium due to grain growth. In
particular, the process occurs in nuclear fuel (e.g.,
UO$_2$),whereupon the fission product gases (e.g., Xe and Kr)
diffusing in the grain can be subjected to a moving grain
boundary. Grain boundary movement can sweep up fission gas atoms more
rapidly than they could have arrived at the boundary by diffusion
\cite{Hargreaves_Collins_1976}. This problem has been treated in
\cite{Forsberg_Massih_2007}. Modelling grain growth phenomenon or
evolution of cellular structure in general has been a subject of many
recent investigations using various theoretical and numerical
approaches, see review by \citeasnoun{C_V_Thompson_2001}.  For
example, in \cite{Niemiec_et_al_1998,Godanski_et_al_1998} a different
approach is used to describe the grain growth process and for which
the exponents of the growth asymptotes are presented.

In this note, we solve the time-dependent diffusion equation for
particles in a $d$-dimensional expanding spherical grain to obtain the
particle density as a function of space and time. Equations of motion
are treated in section \ref{sec:gov-eqs}. General solutions for
particle density are provided in section \ref{sec:general}. Special
solutions are derived in section \ref{sec:special} for the case of
$d=3$ and a parabolic grain growth. An evaluation of particle survival
is presented in section \ref{sec:evaluate}; here we also include
asymptotic solutions for particle density when grain growth rate is
small relative to particle diffusivity and vice versa. We conclude the
note in section \ref{sec:summary} with a brief summary of the results.

\section{Equations of motion}
\label{sec:gov-eqs}

We consider the equivalent sphere model for material, meaning that the
grains of the material are treated as a collection of spheres of
uniform size characterized by a single equivalent radius, $R \equiv
R(t)$, which is a function of time $t$. The particles
(e.g. interstitials, impurities etc) migrate to grain boundaries by
diffusion upon which are released from the system. The grain can
contain traps (vacancies, pores, etc), which may capture the diffusing
particle during its flight. The differential equation for the density
of particles at position $r$ in a $d$-dimensional spherical grain at
time $t$, $C(r,t)$, is given by the law of isotropic diffusion, viz.
 \begin{equation}
 \frac{\partial C(r,t)}{\partial t} = D\nabla_r^2 C(r,t),
 \label{eqn:difeq1}
 \end{equation}
\noindent
subject to the conditions,
 \begin{eqnarray}
\label{eqn:bc1}
 \frac{\partial C(0,t)}{\partial r} & = & 0, \\
\label{eqn:bc2}
 C(R(t),t) & = & 0,\\
\label{eqn:bc3}
 C(r,0) & = & C_0,
 \end{eqnarray}
\noindent
where $D$ is the effective diffusivity of particles in the grain,
accounting for the presence of traps, $\nabla_r^2=\partial^2/\partial
r^2+((d-1)/r)\partial/\partial r $, $C_0$ is the initial
density of particles in the grain, and equation (\ref{eqn:bc2})
defines the absorbing boundary (Dirichlet boundary condition). We should note that $D$, the
effective diffusion coefficient of the particles in the grain, accounts for the
effect of intragranular traps. As has been shown by
\citeasnoun{Speight_1969}, for stationary traps, it is related to the intrinsic diffusivity
$D_{int}$ of particle in the solid through $D= p D_{int}/(p+g)$,
where $p$ is the escape probability of the trapped particle and $g$
is the corresponding capture probability, related to trap size, the
number of traps $n$ and the intrinsic diffusivity, e.g. for spherical
traps, $g=4\pi D_{int}\varrho n$, where $\varrho$ is the trap mean
radius \cite{Ham_1958}. Thus at any time, only a fraction of the
particles is trapped, while the rest are able to
diffuse out of the grain.

The total number of particles per unit volume of $d$-dimensional
spherical grain ($V_d$) at time $t$, $G(t)$, is
given by
\begin{equation}
\label{eqn:inteq1}
 G(t)  =  \frac{d}{R^d(t)} \int_{0}^{R} r^{d-1}C(r,t)\,dr.
\end{equation}
\noindent
The total number of particles contained in the grain at time $t$ is simply
$S(t)\equiv V_d(t)G(t)$ with $V_d(t)=\pi^{d/2}R^d(t)/\Gamma(d/2+1)$
being the volume of sphere at time $t$; while the total number of particles released from
the grain is
\begin{equation}
\label{eqn:inteq2}
 F(t)  =  V_d(0)C_0-S(t).
 \end{equation}
\noindent
 In general, we allow the particle diffusion and grain growth to occur
 simultaneously; hence we scale equation (\ref{eqn:difeq1}) to embed
the time-dependent variable $R(t)$ in the coordinate of the
partial differential equation (space-time), namely
\begin{equation}
\frac{\partial \mathcal{C}(\rho,\tau)}{\partial \tau}  = 
 \frac{\partial^2 \mathcal{C}(\rho,\tau)}{\partial \rho^2} + 
 \Big[\frac{d-1}{\rho}+\frac{R_\tau}{R}\rho\Big]
\frac{\partial \mathcal{C}(\rho,\tau)}{\partial \rho}, 
 \label{eqn:difeq2}
 \end{equation}
\noindent
where we put
 \begin{eqnarray}
\label{eqn:rho}
 \rho & = & \frac{r}{R}, \\
\label{eqn:tau}
 \tau & = & D \int_{0}^{t}\frac{ds}{R^2(s)},
 \end{eqnarray}
\noindent
with $R_\tau=\partial R/\partial\tau$ and the conditions:  $\partial\mathcal{C}(0,\tau)/\partial \rho=0$,
$\mathcal{C}(1,\tau)=0$, and $\mathcal{C}(\rho,0)=C_0$.
The total number of particles contained in the grain as a function of
$\tau$  is obtained by transforming  equation (\ref{eqn:inteq1}) to
\begin{equation}
\label{eqn:varinteq2}
 \mathcal{G}(\tau)= d\int_{0}^{1}x^{d-1}\mathcal{C}(x,\tau)dx.
 \end{equation}
\noindent
Note that in our convention the variables are transformed to
$C(r,t)\Rightarrow \mathcal{C}(\rho,\tau)$, $G(t)\Rightarrow
\mathcal{G}(\tau)$ and $R(t(\tau)) = \mathcal{R}(\tau)$. Solving
equation (\ref{eqn:difeq2}) for $ \mathcal{C}(\rho,\tau)$,
$\mathcal{G}(\tau)$ can be determined through the evaluation of the
integral in (\ref{eqn:varinteq2}).

Before discussing the solution of equation (\ref{eqn:difeq2}), we
should note that grain growth is assumed to obey a power law
\cite{Atkinson_1988,Martin_et_al_1997} according to
\begin{equation}
\label{eqn:gglaw}
 \bar{R}(t)^m -\bar{R}_0^m = Kt,
\end{equation}
\noindent
where $\bar{R}$ is the mean grain radius and $K$ is a constant with
Arrhenius temperature dependence. In the limit where $\bar{R}(t)\gg
\bar{R}_0$, we write $\bar{R}(t) = (Kt)^\alpha$, with $\alpha\equiv
1/m$ being the grain growth exponent. In mean field (non-topological)
theories $\alpha=1/2$ for both $d=2$ and $d=3$, i.e. a parabolic growth
law. Grain growth experiments usually exhibit $\alpha \le 1/2$, the
difference could be a result of neglecting the topological constraints
and the detailed microstructure.

\section{General solution}
\label{sec:general}

Following the method outlined in \cite{Forsberg_Massih_2007}, we write the
general solution of equation (\ref{eqn:difeq2}) in terms of the basis
components
\begin{equation}
\label{eqn:homsol}
 \mathcal{C}(\rho,\tau) = \sum_{i=0}^\infty \mathfrak{a}_i \, e_i(\rho,\tau),
  \end{equation}
\noindent
where $\mathfrak{a}_i$ are real numbers (constants). Next, we separate
the principal time-dependence of $e_i(\rho,\tau)$ by making an ansatz
\begin{equation}
\label{eqn:ansatz}
  e_i(\rho,\tau) = \exp{(-\omega_i \tau})\, \widetilde{e}_i(\rho,\tau),
 \end{equation}
\noindent
where  $\omega_i$ is a constant and $\widetilde{e}_i$ is weakly time-dependent obeying
\begin{eqnarray}
 \label{eqn:homdifeq-2}
 \frac{\partial {\widetilde{e}_i}}{\partial \tau}   = 
 \triangle_\rho'\,{\widetilde{e}_i} +\omega_i \, {\widetilde{e}_i}\\
\label{eqn:hominteq-2}
   \textrm{with} \qquad \triangle_\rho' \equiv \frac{\partial^2 }{\partial \rho^2} + 
 \Big[\frac{d-1}{\rho}+\frac{R_\tau}{R}\rho\Big]
\frac{\partial}{\partial \rho},
  \end{eqnarray}
\noindent
which is subjected to the boundary conditions:
$\partial{\widetilde{e}_i}(0,\tau)/\partial \rho=0$ and
${\widetilde{e}_i}(1,\tau)=0$.

 We now express the function $\widetilde{e}_i$ in terms of an infinite
series

\begin{equation}
\label{eqn:homsol-2}
   \widetilde{e}_i = \sum_{k=0}^\infty \, {_i§\lambda_k} \, {_k§ \widehat{e}_i},
  \end{equation}
\noindent
where $_i§\lambda_k; \: i \ge 1$, with $_i§\lambda_0=1$, and
$\omega_i$ are selected such that the imposed boundary
conditions are satisfied for all $\tau$ and $_k§ \widehat{e}_i$
satisfy  equation (\ref{eqn:homdifeq-2}) with solutions expressed in binomial series in
the form:
\begin{equation}
\label{eqn:homsol-3}
  {_k§ \widehat{e}_i}(\rho,\tau)   = \sum_{m=0}^k {k \choose m} (2\tau)^{k-m} \: {_i^m\!f}(\rho),
 \end{equation}
\noindent
where $_i^m\!f$ satisfy the following differential equations
\begin{eqnarray}
\label{eqn:homdifeq-3}
 \qquad \triangle_\rho'\;{_i^0\!f} +
 \omega_i\;{_i^0\!f}  & = & 0;  \\
\label{eqn:homdifeq-4}
 \qquad \triangle_\rho'\;{_i^m\!f} +
 \omega_i\;{_i^m\!f}  & = & 2m\;{_i^{m-1}\!f};  \quad \textrm{for} \:\:  m\ge 1
  \end{eqnarray}
\noindent
with  ${_i^m\!f}(1)=0$, ${_i^m\!f}(0)=\delta_{m0}$ and ${_i^m\!f}_\rho(0)=0$; 
$\delta_{m0}$ the Kronecker delta and $f_\rho=df/d\rho$. As will be shown
below, ${_i^0\!f}$ are expressible in terms of the confluent
hypergeometric functions, whereas ${_i^m\!f}$ with $m\ge 1$ are
related to such functions. 

Writing equation (\ref{eqn:homdifeq-3}) in a spherical grain, we obtain
\begin{equation}
\label{eqn:homspheq}
 \frac{d^2\mathbf{f}}{d^2\rho} + \big(\frac{d-1}{\rho}+\mathbf{c}\rho\big)\frac{d\mathbf{f}}{d\rho} +
 \mathbf{w}\mathbf{f}=0,
  \end{equation}
\noindent
subject to $\mathbf{f}(1)=0$, $\mathbf{f}(0)=1$ and
$\mathbf{f}_\rho(0)=0$, where $\mathbf{f}={_i^m\!f}$,
$\mathbf{c}=\partial\ln\mathcal{R}/\partial
 \tau\equiv\mathcal{R}_\tau/\mathcal{R}$ and $\mathbf{w}=\omega_i$. Equation (\ref{eqn:homspheq}) can be
transformed to the more familiar confluent hypergeometric equation
\begin{equation}
\label{eqn:kummer-eq1}
 z\frac{d^2\mathbf{u}}{d^2z} + \big(\frac{d}{2}-z\big)\frac{d\mathbf{u}}{dz}
 - \mathbf{a}\mathbf{u}=0,
  \end{equation}
\noindent
subject to $\mathbf{u}(-c/2)=0$, $\mathbf{u}(0)=1$ and that $\mathbf{u}_z(0)$ exists, where
$z=-\mathbf{c}\rho^2/2$, $\mathbf{a}=\mathbf{w}/2\mathbf{c}$ and
$\mathbf{u}(z)\equiv\mathbf{f}(\rho)$. We should note that equations
(\ref{eqn:homspheq}) and (\ref{eqn:kummer-eq1}) are tacitly
time-dependent through $\mathbf{c}$. When
$\mathcal{R}_\tau/\mathcal{R}=$constant, $\mathbf{c}$ becomes
a time-independent constant and exact solutions can be derived. 
The assumption that $\mathbf{c}$= constant is exact for a parabolic growth law, since by
recalling equation (\ref{eqn:tau}), it implies $dR^2/dt=2D\mathbf{c}$.

A solution of equation (\ref{eqn:kummer-eq1}) is the confluent
hypergeometric function \cite{Dennery_Krzywicki_1995} written in
the form  
\begin{equation}
\label{eqn:kummer-eq1-sol}
 \mathbf{u}(z)=  \Phi\big(\mathbf{a},\frac{d}{2};z\big),
  \end{equation}
\noindent
where $\Phi(a,b;z)$ is the Kummer function; also with alternative
notations $M(a,b,z)$ or $_1\!F_1(a,b,z)$, see e.g.,
\cite{Abramowitz_Stegun_1964}. This function is regular at $z=0$ and
can be expressed as a power series:
\begin{equation}
\label{eqn:kummer-eq1-ser}
  \Phi(a,b;z)=
\sum_{s=0}^\infty\frac{\Gamma(a+s)\Gamma(b)z^s}{\Gamma(a)\Gamma(b+s)\Gamma(1+s)}.
 \end{equation}

\section{Special solutions}
\label{sec:special}
 
A point worth noting is that equation (\ref{eqn:kummer-eq1}) can be
expressed as an eigenvalue problem of the form
$\mathcal{L}u(z)=a_nu(z)$, where $a_n$ is the energy eigenvalue of
the operator $\mathcal{L}=zd^2/dz^2+(d/2-z)d/dz$. We ``quantize'' the
eigenvalues: $a_n=n+1/2$, where $n=0,1,2,\dots$ ($n\in
\mathbb{N}$) and thus write
\begin{equation}
\label{eqn:kummer-eq1-sol2}
 {_n\!u}(z)=  \Phi\big(n+\frac{1}{2},\frac{d}{2};z\big),
  \end{equation}
or alternatively
\begin{equation}
\label{eqn:kummer-eq1-sol3}
 {_n^0\!f}(y)=  \Phi\big(n+\frac{1}{2},\frac{d}{2};-y^2\big).
  \end{equation}
Here the choice of $a_n = n+1/2$ is merely to obtain a series of
convenient expressions (harmonics) for the solutions ${_n^0\!f}(y)$
and integrals of ${_n^1\!f}(y)$, see below.  For the ground state, $n=0$ and  $d=3$, we have
\begin{equation}
\label{eqn:kummer-eq1-sol4}
 {_0^0\!f}(y)=  \frac{\sqrt{\pi}}{2y}\;\mathrm{erf}(y),
  \end{equation}
 where $y=(c/2)^{1/2}\rho$ and $\mathrm{erf}(y)$ is the usual error
function.

 Next, returning to equation (\ref{eqn:homdifeq-4}) and writing it in the spherical system,

\begin{equation}
\label{eqn:homspheq-2}
 {_i^m\!f}_{\rho\rho} + \Big(\frac{d-1}{\rho}+c\rho\Big)\;{_i^m\!f}_{\rho}+ \omega_i\; {_i^m\!f} 
=  2m \;{_i^{m-1}\!f};  \quad \textrm{for} \:\:  m\ge 1
 \end{equation}
Again, replacing $z=-c\rho^2/2$, equation (\ref{eqn:homspheq-2}) is
transformed to 
\begin{equation}
\label{eqn:homspheq-3}
 z\;{_n^m\!u}_{zz} + \Big(m+\frac{1}{2}-z\Big)\;{_n^m\!u}_{z}- \Big(n+\frac{1}{2} \Big)\; {_n^m\!u} 
=  -\frac{m}{c} \;{_n^{m-1}\!u};  \quad \textrm{for} \:\:  m\ge 1
 \end{equation}
where the aforementioned quantized energy eigenvalues, $\omega/2c=n+1/2$, were
used, and in addition, the spatial dimension was quantized, in the
manner, $d=2m+1$. Let us write (\ref{eqn:homspheq-3}) for $m=1$ and $n\in
\mathbb{N}$ with the convention ${_n\!u}\equiv {_n^1\!u}$, namely
\begin{equation}
\label{eqn:homspheq-4}
 z\;{_n\!u}_{zz} + \Big(\frac{3}{2}-z\Big)\;{_n\!u}_{z}- \Big(n+\frac{1}{2} \Big)\; {_n\!u} 
=  -\frac{{_n\!p}(z)}{c}\; e^z.
 \end{equation}
In writing the right-hand side of equation (\ref{eqn:homspheq-4}), we
have utilized Kummer's relation
\begin{eqnarray}
\label{eqn:kummer-id1}
 \Phi\big(n+\frac{1}{2},m+\frac{1}{2};z\big) =  e^z\;{_n^m\!p}(z),\\
\label{eqn:kummer-id2}
 {_n^m\!p}(z) =  \Phi\big(m-n,m+\frac{1}{2};-z\big),
 \end{eqnarray} 
where ${_n^m\!p}(z)$ is a polynomial when $n\ge m$. In particular, with
${_n\!p}={_n^1\!p}$, we have
\begin{eqnarray}
\label{eqn:kummer-poly1}
 {_1\,p}(z) & = & 1,\\
\label{eqn:kummer-poly2}
{_2\,p}(z) & = & 1+\frac{2z}{3},\\
\label{eqn:kummer-poly3}
 {_3\,p}(z) & = & 1+\frac{4z}{3}+\frac{4z^2}{15},\\
\label{eqn:kummer-poly4}
 {_4\,p}(z) & = & 1+2z+\frac{4z^2}{5}+\frac{8z^3}{105},
 \end{eqnarray} 
Let us now remove $e^z$ from equation (\ref{eqn:homspheq-4}) by
making the substitution ${_n\!u}(z)=e^z{_n\!v}(z)$, we obtain
\begin{equation}
\label{eqn:homspheq-5}
 z\;{_n\!v}_{zz} + \big(\frac{3}{2}+z\big)\;{_n\!v}_{z}- \big(n-1 \big)\; {_n\!v} 
=  -\frac{{_n\!p}(z)}{c}.
 \end{equation}
This can be differentiated $n-1$ times to give 
\begin{equation}
\label{eqn:homspheq-n-1}
 z\;{_n\!v}^{(n+1)} + \big(n+\frac{1}{2}+z\big)\;{_n\!v}^{(n)} =  -\frac{2^{n-1}(n-1)!}{(2n-1)!!c},
 \end{equation}
where the superscript in the parenthesis on ${_n\!v}$ denotes the number of
differentiations with respect to the argument.  We find
${_n\!v}$ by first integrating (\ref{eqn:homspheq-n-1}) once, which yields
\begin{equation}
\label{eqn:integsol-n}
 {_n\!v}^{(n)}(z) =
-\frac{2^{n-1}(n-1)!}{(2n-1)!!c}z^{-(n+1/2)}e^{-z}\int_0^z \zeta
^{n-1/2}e^\zeta d\zeta,
 \end{equation}
then with further integrations, $n$ times, we obtain
\begin{equation}
\label{eqn:intgsol-1a}
 {_n\!v}(z) =
-\frac{2^{n-1}}{(2n-1)!!c}\int_0^z (z-s)^{n-1}\int_0^s
 \Big(\frac{\zeta}{s}\Big)^{n-1/2} e^{\zeta-s}\; d\zeta ds.
 \end{equation}
After some rearrangement, we write
\begin{equation}
\label{eqn:intgsol-1b}
 {_n\!v}(z) =
-\frac{2^{n-1}}{(2n-1)!!c}\int_0^1 (1-\kappa)^{n-1/2}e^{-\kappa z}\int_0^z
 s^{n-1} e^{\kappa s}\; ds d\kappa.
 \end{equation}
The integral over the variable $z$ can be expressed in terms of gamma
functions, viz.
\begin{equation}
\label{eqn:intgsol-1c}
 \int_0^z s^{n} e^{\kappa s}\; ds =
(-\kappa)^{-n-1}\big[\Gamma(n+1,-\kappa z)-n\Gamma(n)\big],
 \end{equation}
where $\Gamma(a,x)$ is an incomplete gamma function
\cite{Abramowitz_Stegun_1964}. 
Substituting this result in equation (\ref{eqn:intgsol-1b}) and making
use of the properties of gamma functions \cite{Abramowitz_Stegun_1964},
we finally write
\begin{equation}
\label{eqn:intgsol-1d}
 {_n\!v}(z) =
-\frac{2^{n-1}(n-1)!}{(2n-1)!!c}\int_0^1 (1-\kappa)^{n-1/2}(-\kappa)^{-n-2}\sum_{j=0}^\infty
 \frac{(-\kappa z)^{n+j}}{\Gamma(n+j+1)} d\kappa.
 \end{equation}
Hence solution to equation (\ref{eqn:homspheq-4}) is found through
${_n\!u}(z)=e^z{_n\!v}(z)$. 

Let us evaluate the solution of equation (\ref{eqn:kummer-eq1}),
relation (\ref{eqn:kummer-eq1-sol}), when $c$ is small, i.e.,
$\mathcal{R}_\tau/\mathcal{R} \ll 1$ or $a \gg 1$. We first write
(\ref{eqn:kummer-eq1-sol}) in a quantized form
\begin{equation}
\label{eqn:kummer-inf-sol}
 \Phi\big(a,\frac{d}{2};z\big) \Rightarrow \Phi\big(a+m,\frac{d}{2}+m;z\big),
  \end{equation}
\noindent
 for $m\in \mathbb{N}$. For negative $z$, we can expand
(\ref{eqn:kummer-inf-sol}) in series
\cite{Abramowitz_Stegun_1964} according to

\begin{equation}
\label{eqn:kummer-expand}
 \Phi\big(a+m,\frac{d}{2}+m;z\big)=
\Gamma\big(m+\frac{d}{2}\big)e^{z/2}\Big(\frac{\sigma}{2}\Big)^\nu
 \sum_{n=0}^\infty A_n\Big(\frac{z}{\sigma}\Big)^{n} J_{n-\nu}(\sigma),
  \end{equation}
\noindent
where $\sigma=\sqrt{(d-2m)z-4az}$, $\nu = 1-m-d/2$, $A_0=1$, $A_1=0$,
$A_2=d/4+m/2$, and

\begin{equation}
\label{eqn:kummer-recurs}
 (n+1)A_{n+1}= (n+d/2+m-1)A_{n-1}+(2a+m-d/2)A_{n-2}.
  \end{equation}
\noindent
In equation (\ref{eqn:kummer-expand}), $J_p(\sigma)$ is the Bessel
function of the first kind, related to the spherical Bessel function via
 $j_p(\sigma)=\sqrt{\pi /2\sigma}J_{p+1/2}(\sigma)$. Specializing to
$d=3$, equation (\ref{eqn:kummer-expand}) can be expressed as
\begin{equation}
\label{eqn:kummer-ser-d3}
 \Phi\big(a+m,\frac{3}{2}+m;z\big)=
(2m+1)!!\;e^{z/2} \sum_{n=0}^\infty A_n z^n \sigma^{-(m+n)} j_{m+n}(\sigma),
  \end{equation}
\noindent
Since $A_1=0$, the first term in series (\ref{eqn:kummer-ser-d3})
offers a good approximation for large $a$. For $m=0$, we find
\begin{equation}
\label{eqn:kummer-ser-d3-1}
 \Phi\big(a,\frac{3}{2};z\big) \approx
 e^{z/2}\; \frac{\sin{\sigma}}{\sigma},
  \end{equation}
\noindent
with $\sigma=\sqrt{(3-4a)z}$. Recalling the substitutions
$a=\omega/2c$ and $z=-c\rho^2/2$, we write
\begin{equation}
\label{eqn:kummer-ser-d3-2}
 \Phi\big(\frac{\omega}{2c},\frac{3}{2};-c\rho^2/2\big) \approx
 \frac{e^{-c\rho^2/4}\sin\big[{\rho\sqrt{\omega-3c/2}}\big]}{\rho\sqrt{\omega-3c/2}}.
 \end{equation}
\noindent
Hence for $c=0$, we have
$\Phi=\sin{(\rho\sqrt{\omega})}/(\rho\sqrt{\omega})$;
the result which could be obtained directly from
equation (\ref{eqn:kummer-eq1-sol}), viz.
\begin{equation}
\label{eqn:kummer-lim}
 \lim_{c \to
0}\Phi\big(\frac{\omega}{2c},\frac{d}{2};-c\rho^2/2\big)=_0\!F_1(;\frac{d}{2};-\omega\rho^2/4).
  \end{equation}
\noindent
The $_0\!F_1$  function has the series expansion
$_0\!F_1(;b;z)=\sum_{k=0}^\infty (\Gamma(b)/\Gamma(b+k))z^k/k!$, and can be expressed
in terms of the Bessel functions of the first kind, e.g.,
$_0\!F_1(;3/2;-\omega\rho^2/4)=j_0(\rho\sqrt{\omega})=\sin{(\rho\sqrt{\omega})}/(\rho\sqrt{\omega})$.

Let us now be more explicit and write the series solution from
equations (\ref{eqn:ansatz}) and (\ref{eqn:homsol-2}) as follows
\begin{equation}
\label{eqn:serie-1}
  e_i(\rho,\tau) =  \Big(\, {_0§ \widehat{e}_i}+ \, {_i§\lambda_1} \, {_1§
\widehat{e}_i} + \, {_i§\lambda_2} \, {_2§ \widehat{e}_i}+\cdots\Big)\exp(-\omega_i\tau),
  \end{equation}
\noindent
where, according to equation (\ref{eqn:homsol-3}), we write
\begin{eqnarray}
\label{eqn:fser}
  \, {_0§ \widehat{e}_i} & = & \: {_i^0\!f}(\rho)\nonumber\\
  \, {_1§ \widehat{e}_i} & = & \: {_i^1\!f}(\rho) + 2\tau \:
  {_i^0\!f}(\rho)\nonumber\\
  \, {_2§ \widehat{e}_i} & = & \: {_i^2\!f}(\rho) + 4\tau\:
{_i^1\!f}(\rho) + 4\tau^2 \: {_i^0\!f}(\rho)\nonumber\\ 
 \textrm{etc}. & 
 \end{eqnarray}
 The functions $_i^m\!f$ satisfy the differential equations
(\ref{eqn:homdifeq-3})-(\ref{eqn:homdifeq-4}) or
(\ref{eqn:homspheq-2}). For $c=0$, i.e. when $(dR^2/dt)/D\ll 1$, we have ($d=3$)
\begin{eqnarray}
\label{eqn:fsolm}
 \: {_i^0\!f}(\rho) & = & j_0(\rho\mu_i) \\
 \: {_i^m\!f}(\rho) & = & \mu_i^{-2m}(\rho\mu_i)^mj_m(\rho\mu_i),
 \end{eqnarray}
where $\mu_i=\sqrt{\omega_i}$. Satisfying the boundary condition $\:
{_i^0\!f}(1)=0$, we obtain 

\begin{equation}
\label{eqn:muj}
 \mu_j=j\pi \qquad \qquad \qquad j=0,1,2, \dots.
 \end{equation}
\noindent

Next, we calculate the the total number of particles per unit volume
of the grain at ``time'' $\tau$; i.e. equation (\ref{eqn:varinteq2}), expressed as

\begin{equation}
\label{eqn:varinteq2-ser1}
 \mathcal{G}(\tau)= \sum_{i=0}^\infty \mathcal{G}_i(\tau),
 \end{equation}
\noindent
where using equation (\ref{eqn:homsol}) and in three dimensions, we write
\begin{equation}
\label{eqn:varinteq2-ser2}
 \mathcal{G}_i(\tau)= 3\int_{0}^{1}x^2e_i(x,\tau)dx.
 \end{equation}
\noindent
Substituting for $e_i(x,\tau)$ from equation (\ref{eqn:serie-1}) and
employing the relations in (\ref{eqn:fser}) gives

\begin{eqnarray}
\label{eqn:varinteq2-ser3}
 \mathcal{G}_i(\tau) & = & 3\int_{0}^{1}x^2\Big[ \: {_i^0\!f}(x)+
\,{_i§\lambda_1}\Big(\: {_i^1\!f}(x) + 2\tau \:
{_i^0\!f}(x)\Big)+\nonumber\\
& & + \,{_i§\lambda_2}\Big(\: {_i^2\!f}(x) + 4\tau\:
{_i^1\!f}(x) + 4\tau^2 \: {_i^0\!f}(x)\Big)+\cdots\Big]e^{-\mu_i^2\tau}dx.
 \end{eqnarray}
\noindent
The terms in equation (\ref{eqn:varinteq2-ser3}) can be integrated by
using the relation:
\begin{equation}
\label{eqn:trig-integral}
   \int_{0}^{1}x^2\;{_i^n\!f}(x)dx = \mu_i^{-1} j_{n+1}(\mu_i).
 \end{equation}
\noindent
In a numerical treatment the parameters $\, {_i§\lambda_1}$,$\,
{_i§\lambda_2}$ are determined by satisfying the boundary conditions at the
beginning and the end of each discretized time interval. Moreover, the
coefficients $\mathfrak{a}_i$ in equation (\ref{eqn:homsol}) are
determined from the the spatial initial condition
$\mathcal{C}(\rho,0)=C_0$. Assuming the orthonormality of the
solutions, then in general

\begin{equation}
\label{eqn:trig-integral}
   \mathfrak{a}_i=\frac{\int_{0}^{1}\mathcal{C}(\rho,0)\;e_i(\rho,0)\rho^{d-1}d\rho}
{\int_{0}^1 e_i(\rho,0)^2\rho^{d-1}d\rho}.
 \end{equation}
\noindent

\section{Evaluation}
\label{sec:evaluate}

Let us now evaluate the time variation of the number of particles
contained in the grain (survival rate), viz.

\begin{equation}
\label{eqn:survive-rate_1}
 \frac{dS}{dt}  =  K_d \frac{\partial}{\partial t}\int_{0}^{R(t)} r^{d-1}C(r,t)dr,
\end{equation}
\noindent
where $K_d=2\pi^{d/2}/\Gamma(d/2)$. Differentiating under the integral and using
equations (\ref{eqn:bc1}) and (\ref{eqn:difeq1}), we obtain
\begin{equation}
\label{eqn:survive-rate_2}
 \frac{dS}{dt}  =  K_d D R(t)^{d-1}\Big[\frac{\partial
C(r,t)}{\partial r}\Big]_{r=R(t)}.
\end{equation}
\noindent
We can also calculate the time derivative of $G(t)$, expressed in the form
\begin{equation}
\label{eqn:G-rate_1}
 \dot{G}(t)  =  d\Big[\frac{\dot{S}(t)}{K_dR(t)^d}-\frac{\dot{R(t)}}{R(t)}G(t)\Big],
\end{equation}
\noindent
where over-dot denotes temporal differentiation. Substituting for
$\dot{S}$ from equation (\ref{eqn:survive-rate_2}) and simplifying, we
find
\begin{equation}
\label{eqn:G-rate_2}
 \dot{G}(t)  =  d\Big[\frac{D}{R}\Big(\frac{\partial C}{\partial r}\Big)_{R}-\frac{\dot{R}}{R}G(t)\Big].
\end{equation}
\noindent
This equation shows the balance between particle diffusion and grain
growth. If we designate the characteristic frequency of particle
diffusion by $\nu_D=D/R^2$ and that of grain growth by
$\nu_G=\dot{R}/R$, we notice that when $\nu_D\ll\nu_G$ the grain
growth is dominating and particles cannot reach the boundary of the
cell and $G=C_0(R_0/R)^d$ with $R_0$ denoting the initial cell
size. Using the relation for the grain growth, $R=(Kt)^\alpha$, we can
write
\begin{equation}
\label{eqn:G_lim1}
 G(t)  \sim t^{-\alpha d} \quad  \textrm{for} \quad \nu_G\gg \nu_D.
\end{equation}
\noindent
For example, in 3-dimensional spherical grain with $\alpha=1/2$,
$G(t)\sim t^{-3/2}$. Thus the total number of particles in the grain
remains constant with time, i.e., $S=V(t)G(t)=$const.

On the other hand when $\nu_D\gg\nu_G$, the diffusion flux is the controlling
parameter for escape rate. It may be illustrating to evaluate
$\dot{G}(t)$ in this regime for $d=3$ when $R$ is stationary,
with the well-known exact solution \cite{Carslaw_Jaeger_1959}
\begin{eqnarray}
\label{eqn:G-rate_3}
 \dot{G}(t) &  = & 3\frac{D}{R}\Big(\frac{\partial C}{\partial
r}\Big)_{R} \quad   \textrm{for} \quad \nu_G\ll 1 \nonumber\\ 
  G(\mathcal{T})  & =  & 6 C_0 \sum_{k=1}^\infty\frac{\exp(-k^2\pi^2\mathcal{T})}{k^2\pi^2}.
\end{eqnarray}
\noindent
Here $\mathcal{T}=Dt/R^2$ and also equation (\ref{eqn:G-rate_3}) gives
$G(0)=C_0$, as it should. In the long-time limit only the slowest
decaying eigenmode contributes; hence the asymptotic number of
particles per unit volume decays according to

\begin{equation}
\label{eqn:G_lim2}
 G(\mathcal{T})  \sim C_0 e^{-\pi^2 \mathcal{T}} \quad  \textrm{for} \quad
\nu_G\ll 1  \; \land \; \mathcal{T} \gg 1.
\end{equation}
\noindent
The corresponding formula for particle survival in the grain is
\begin{equation}
\label{eqn:S_lim2}
 S(t)  \sim (Kt)^{3\alpha} e^{-\pi^2 D K^{-2\alpha} t^{1-2\alpha}} \quad  \textrm{for} \quad
\nu_G\ll 1  \; \land \; Dt/R^2 \gg 1,
\end{equation}
\noindent
where again we employed $R=(Kt)^\alpha$. Hence for $\alpha < 2$, $S(t)$
exponentially decays to zero at long times.

Let us finally investigate the case of a slowly growing grain by means of the
adiabatic approximation \cite{Krapivsky_Redner_1996}. In this method,
the asymptotic solution is constructed to have the same functional
form as in the stationary grain, but in addition to satisfy the
time-dependent boundary condition, namely, $ C(R(t),t) = 0$ and
$C(r,0)=C_0$. We write ($d=3$)

\begin{equation}
\label{eqn:ad-ap}
 C(r,t) \approx \frac{R(t)}{r}\cos\Big(\frac{\pi
r}{2R(t)}\Big)f(t)\equiv C_{ad}(r,t),
\end{equation}
\noindent
where the function $f(t)$ is determined by substituting equation
(\ref{eqn:ad-ap}) into the diffusion equation (\ref{eqn:difeq1}),
which yields
\begin{equation}
\label{eqn:fad-ap1}
 \dot{f}(t) = -\frac{\pi^2D+4R\dot{R}}{4R^2}\Big[1+\frac{2\pi
r\dot{R}}{\pi^2D+4R\dot{R}}\tan\Big(\frac{\pi
r}{2R}\Big)\Big]f(t).
\end{equation}
\noindent
Now by placing  $R=(Kt)^\alpha$, with $\alpha < 1/2$, into equation
(\ref{eqn:fad-ap1}) the second term in the square bracket is negligible and we write
\begin{equation}
\label{eqn:fad-ap2}
 \frac{df}{dt} \approx -\big(\frac{\pi^2D}{4R^2}+\frac{\dot{R}}{R}\big)f,
\end{equation}
\noindent
If we further consider that  $\nu_D\gg\nu_G$, then $f(t)$ is
given by
\begin{equation}
\label{eqn:fad-sol}
 f(t) \approx \exp\Big[-\frac{\pi^2D}{4(1-2\alpha)K^{2\alpha}}t^{1-2\alpha}\Big],
\end{equation}
\noindent
which is the same  result obtained by  \citeasnoun{Krapivsky_Redner_1996} for the
one-dimensional case. The corresponding relation  for $S(t)$ is
\begin{equation}
\label{eqn:S_adap}
 S(t)  \approx K^{3\alpha}t^{3\alpha}
\exp\Big[-\frac{\pi^2D}{4(1-2\alpha)K^{2\alpha}}t^{1-2\alpha}\Big],
\end{equation}
\noindent
which is consistent with equation (\ref{eqn:S_lim2}).
\section{Summary}
\label{sec:summary}

In this note we have treated the problem of particles undergoing
Brownian motion in an expanding sphere whose surface absorb the
particles (Dirichlet condition). The time-dependent diffusion equation
for particles in $d$-dimensional sphere, which accounts explicitly for
its expansion is formulated. The equation is solved by separating the
space and time variables and the spatial equation for the particle
density is reduced to an ordinary confluent hypergeometric
differential equation, which implicitly contains the grain growth
process. For a parabolic grain growth law analytical solutions in
terms of special functions are derived. When grain growth is dominant
over particle diffusion, the number of particles per unit volume
contained in the grain is calculated to evolve with time as $G(t) \sim
t^{-\alpha d}$, where $\alpha$ is the growth exponent and $d$ the
spatial dimensionality. For the case of slow growth, i.e. when the
ratio of grain growth rate to diffusivity is small an asymptotic
solution is attained. In this regime the long-time limit gives
$G(t)\sim \exp[-\pi^2(D/K^{2\alpha})t^{1-2\alpha}]$, where $D$ is the
particle diffusivity and $K$ the growth constant. The paper has
focused on the mathematical and physical aspects of the problem,
numerical evaluations for a similar kind of problem are treated
elsewhere \cite{Forsberg_Massih_2007}.


\section*{References}
\bibliographystyle{jphysicsB} \bibliography{frel}

\end{document}